\documentstyle[12pt]{article}

\title{Generic chiral superfield model on nonanticommutative ${\cal N}={1\over 2}$ superspace}

\author{O.D. Azorkina$^{1}$\footnote{azorkina@tspu.edu.ru},
A.T. Banin\footnote{atb@math.nsc.ru}, I.L.
Buchbinder$^{1}$\footnote{joseph@tspu.edu.ru}, N.G.
Pletnev$^{2}$\footnote{pletnev@math.nsc.ru}}

\date{\it
${}^{1}$Department of Theoretical Physics\\
Tomsk State Pedagogical University, Tomsk\\
634041, Russia\\
${}^{2}$Department of Theoretical Physics\\
Institute of Mathematics, Novosibirsk, \\
630090, Russia\\
 }

\begin{document}

\begin{titlepage}

\maketitle

\begin{abstract}
We consider the generic nonanticommutative model of
chiral-antichiral superfields on ${\cal N}={1\over 2}$ superspace.
The model is formulated in terms of an arbitrary K\"ahlerian
potential, chiral and antichiral superpotentials and can include
the nonanticommutative supersymmetric sigma-model as a partial
case. We study a component structure of the model and derive the
component Lagrangian in an explicit form with all auxiliary fields
contributions. We show that the infinite series in the classical
action for generic nonanticommutative model of chiral-antichiral
superfields in $D=4$ dimensions can be resumed in a compact
expression which can be written as a deformation of standard
Zumino's lagrangian and chiral superpotential. Problem of
eliminating the auxiliary fields in the generic model is discussed
and the first perturbative correction to the effective scalar
potential is obtained.
\end{abstract}
\thispagestyle{empty}

\end{titlepage}

Supersymmetric field theories on deformed superspaces with
nonanticommuting coordinates possess the interesting properties in
classical and quantum domains. Remarkable class of such theories
based on special deformation of ${\cal N}=1$ supersymmetry was
proposed by Seiberg \cite{seib}. Seiberg's type of superspace
deformation introduces the nonanticommutativity both even and odd
coordinates but preserves anticommutativity in the chiral sector.
As a result, the corresponding deformed superspace breaks the
supersymmetry in the antichiral sector and therefore it is called
${\cal N}={1\over 2}$ superspace. Formulation of analogous
deformation in ${\cal N}=2$, $D=4$ superspace was given in
\cite{n11st}. Studying of various aspects of ${\cal N}= {1\over
2}$ supersymmetric theories has been carried out in a number of
recent papers (see e.g \cite{berk}, \cite{quantn12}, \cite{renorm}
for $D=4$ models and \cite{D2}, \cite{D24} for $D=2$ models). It
is worth pointing out that the most general deformation of ${\cal
N}=1$ superspace was constructed in work \cite{klemm}.

To interpret the ${\cal N}={1\over 2}$ supersymmetric theories as
conventional field models and to clarify their dynamics it is
necessary to rewrite such superfield theories in the component
form. Finding the component structure of the nonanticommutative
theories is a highly nontrivial technical problem because of the
very complicated superspace structure and therefore it demands a
special study. Component form of actions for
nonanticom\-mu\-tative theories in addition to standard terms
always will contain the terms dependent on the superspace
deformation parameter. Since a half of supersymmetries is broken
down a symmetry between chiral and antichiral superspace
coordinates is absent and some component fields can enter in the
action in very cumbersome combinations.

In the papers \cite{seib},  the component structure of $D=4$,
${\cal N}={1\over 2}$ supersymmetric models Yang-Mills theory and
the Wess-Zumino  was studied. For this case it was shown that the
deformed theory is renormalizable \cite{quantn12}, \cite{renorm}
in spite of the presence of higher dimensional terms in the
Lagrangian and preserves locality. However the generic $D=4$,
${\cal N}={1\over 2}$ supersymmetric chiral-antichiral theories in
four dimensions, which formulated in terms of an arbitrary
K\"ahlerian potential $K(\bar{\Phi},\Phi)$ and arbitrary chiral
and antichiral superpotentials $W(\Phi)$, $\bar{W}(\bar{\Phi})$,
have not been considered in the literature\footnote{Though there
is a well known connection between $D=4$, ${\cal N}=1$ and $D=2$,
${\cal N}=2$ superspaces (for recent review see Refs.
\cite{D2},\cite{D24} and reference therein) a distinct feature
appears in two dimensional models in comparison to four
dimensional models. Such theories in two dimensions appear as a
result of nontrivial reduction from $D=4$ chiral and vector
multiplet. Particularly in addition to the chiral multiplet it is
also possible twisted multiplets. Two-dimensional models have
additional remarkable feature (like conformal and mirror
symmetries) and other special properties. Therefore one can expect
that the various reductions from $D=4$ nonanticommutative generic
chiral-antichiral superfield model to $D=2$ will lead to various
$D=2$ models and structure of $D=4$ model can not be restored on
the base of the known structure of $D=2$ model. This circumstance
justifies necessity for independent study of $D=4$
nonanticommutative models.}. We will call such theories the
generic chiral superfield models. It is worth pointing out that
just generic chiral superfield models (with ${\cal N}=1$
supersymmetry) emerges in the low-energy limit of the superstring
theory (see e.g. \cite{GSW}) and widely used in the superstring
phenomenology (see e.g. \cite{CCE}).

In this paper we study the $D=4$ generic chiral superfield model
in ${\cal N}={1\over 2}$ superspace and derive its component
structure. Explicit expressions for the K\"ahlerian, chiral and
antichiral superpotentials are not fixed. We show that the
component action is represented as an infinite series in
nonanticommutativity parameter with coefficients depending on the
derivatives of above potentials. Despite this fact, it is possible
to write down the action in a closed form via smoothing integrals
of the K\"ahlerian $K$ and chiral $W$ superpotential around the
bosonic component of the chiral superfield $\Phi$ on a scale
dependent on the deformation parameter and the auxiliary field
$\sqrt{\det C}F$. This effect is in an agreement with an
observations of Ref. \cite{alva} for $D=2$. For ${\cal N}=2$ sigma
model nonanticommutativity induces simple deformations of the
Zumino Lagrangian along with the holomorphic superpotential. This
phenomena is interpreted as a fuzziness in target space controlled
by the vacuum expectation value of the auxiliary field.

For $D=4$ sigma-model with a pure K\"ahler potential we
demonstrate that like in $D=2$ sigma-model \cite{chandr} one can
redefine the target space metric by such a way that the symmetry
between the the holomorphic and anti-holomorphic terms in the
action became formally restored. We also discuss a problem of
eliminating the auxiliary fields for the case of constant physical
fields. We think this allows to examinate classical structure of
vacua for the deformed theory.

We begin with consideration of the ${\cal N}= {1\over 2}$ deformed
superspace. According to Seiberg, the coordinates of this
superspace are defined such a way that Grassmannian coordinates
are not complex conjugate to one another ($(\theta^{\alpha})^*
\neq \bar{\theta}^{\dot\alpha}$) and the anticommuting coordinates
$\theta$ form a Clifford algebra
\begin{equation}\label{deform}
\{\theta^{\alpha},\theta^{\beta}\}=
C^{\alpha\beta}~,
\end{equation}
where $C^{\alpha\beta}=C^{\beta\alpha}$ is a symmetrical constant matrix.
The other commutation relations are determined by the consistency
of the algebra:
\begin{equation}\label{alg}
[x^{m},\,\theta^{\alpha}]=i C^{\alpha\beta}\sigma^{m}_{\beta
\dot{\alpha}}\bar{\theta}^{\dot{\alpha}}~,\quad [x^m,\,x^n] =
\bar{\theta}\bar{\theta}\, C^{mn}~,\quad
\{\bar{\theta}^{\dot{\alpha}},\bar{\theta}^{\dot{\beta}}\}=0~,
\end{equation}
where
$C^{mn}=C^{\alpha\beta}\varepsilon_{\beta\gamma}\sigma^{mn\,\gamma}_{~~\alpha}$,
$\bar{\theta}\bar{\theta} = \bar{\theta}_{\dot\alpha}\bar{\theta}^{\dot\alpha}= 2\bar{\theta}^2$.
In contrast to the spacetime coordinates $x^{m}$, the chiral coordinates
$y^{m}=x^{m}+
{i\over 2}\theta^{\alpha}\sigma^{m}_{\alpha\dot{\alpha}}\bar{\theta}^{\dot{\alpha}}$
can be taken commuting, while the antichiral coordinates
$\bar{y}^{m}=y^{m}-
i\theta^{\alpha}\sigma^{m}_{\alpha\dot{\alpha}}\bar{\theta}^{\dot{\alpha}}$ do not commute.
 Because the anticommutation
relation of $\bar{\theta}$ remains undeformed, $\bar{\theta}$ is
not the complex conjugate of $\theta$, that is possible only in
 the Euclidean space. Covariant derivatives and supercharges  are
defined by the standard expressions
\begin{equation}
D_{\alpha} = {\partial\over\partial\theta^{\alpha}} + i\bar{\theta}^{\dot\alpha}\sigma^{m}_{\alpha\dot{\alpha}}{\partial\over\partial y^{m}}~, \quad
\bar{D}_{\dot\alpha} =   {\partial\over\partial\bar{\theta}^{\dot\alpha}}~,
\end{equation}
\begin{equation}
  Q_{\alpha} = i{\partial\over\partial\theta^{\alpha}} ~, \quad
  \bar{Q}_{\dot\alpha} =   i{\partial\over\partial\bar{\theta}^{\dot\alpha}}+\theta^{\alpha}\sigma^{m}_{\alpha\dot{\alpha}}{\partial\over\partial y^{m}}~,
\end{equation}
The generators $Q_{\alpha}$ induce the transformations
corresponding to unbroken ${\cal N}=\frac{1}{2}$ supersymmetry.

Because of the deformation (\ref{deform}) functions of $\theta$
must be ordered. The simplest possible ordering is the Weyl
type\footnote{In used notations $\psi^2= {1\over
2}\psi^{\alpha}\psi_{\alpha}$. } $:\theta^{\alpha}\theta^{\beta}:
\,= {1\over 2}[\theta^{\alpha}, \,\theta^{\beta}] =  -
\varepsilon^{\alpha\beta}\theta^2$. When functions of $\theta$ are
multiplied, the result should be reordered again. Reordering is
implementing by a noncommutative $\star$-product which is defined
as follows
\begin{equation}\label{star:def}
    \Phi\star\Psi = \Phi\,{\rm e}^{\displaystyle - C^{\alpha\beta}\overleftarrow{Q_{\alpha}}\overrightarrow{Q_{\beta}}}\,\Psi
    = \Phi\left(1-C^{\alpha\beta}\overleftarrow{Q_{\alpha}}\overrightarrow{Q_{\beta}} +
    \lambda
    \overleftarrow{Q}^2\overrightarrow{Q}^2\right)\Psi~,
\end{equation}
where $\lambda = -{1\over 2}C^{\alpha\beta}C_{\alpha\beta}$. The
star-product (\ref{star:def}) is invariant under the action of
$Q_{\alpha}$ but is not invariant under action of $\bar{Q}$.

Described deformation preserves the half of ${\cal N}=1$
supersymmetry and has interesting properties in the field theory
viewpoint. Replacing all ordinary products with the above
$\star$-product, one can proceed studying a supersymmetric field
theory in this nonanticommuting superspace taking into account
that this deformed supersymmetry algebra admits well-defined
representations. Namely, the chiral superfield $\Phi$ is defined
by the standard relation $\bar{D}_{\dot{\alpha}}\Phi =0$, which
means $\Phi(y,\theta)= \phi(y)+\theta\kappa(y) -\theta^2 F(y)$.
Antichiral superfield is also defined by the standard relation
$D_{\alpha} \bar{\Phi}=0$, which means
$\bar{\Phi}(\bar{y},\bar{\theta})= \bar{\phi}(\bar{y})
+\bar{\theta}\bar{\kappa}(\bar{y}) -\bar{\theta}^2
\bar{F}(\bar{y})$. As it has been demonstrated in Ref.
\cite{seib}, the $\star$-product (\ref{star:def}) of the chiral
superfields is again a chiral superfield; likewise, the
$\star$-product of the antichiral superfields is again an
antichiral superfield. This observation allows to extend
well-studied anticommutative theories on nonanticommutative
versions by simple replacement the point product with the star
product.

The action of the generic chiral superfield model on ${\cal
N}=1/2$ superspace is \footnote{The integration over Grassmannian
coordinates in normalized as $\int d^2\theta = {1\over
2}{\partial^2\over\partial\theta^2}$, $\int d^2\theta\,\theta^2
=-1$, $\int d^2\bar\theta\,\bar\theta^2 =-1$.}
\begin{equation}\label{genact}
    S_{\star}[\bar{\Phi}, \Phi] = \int d^4x d^4\theta\, K(\bar{\Phi},
    \Phi)_{\star} + \int d^4x d^2\theta\,W(\Phi)_{\star} + \int d^4x
    d^2\bar\theta\,\bar{W}(\bar\Phi)_{\star}~,
\end{equation}
where $K(\bar{\Phi},\Phi), W(\Phi), \bar{W}(\bar{\Phi})$ are the
arbitrary K\"ahlerian potential, chiral and antichiral
superpotentials respectively and the superfield multiplication is
defined in terms of the star-product (\ref{star:def}).

Since the star-product (\ref{star:def}) always begins with the
point product $A\star B= A\cdot B+ \cdots$, it is easy to
understand that the action (\ref{genact}) can be written as a sum
of the action for the general chiral superfield model on
undeformed ${\cal N}=1$ superspace (see for example \cite{idea})
and some contributions higher dimensions resulting from
deformation of the superspace. But the action is local. Further we
will write the action (\ref{genact}) in component fields and study
its structure.

We consider the chiral superpotential component form and write it
as a Taylor series:
\begin{equation}\label{chir}
    \int d^4x d^2\theta\, W(\Phi)_{\star}= \sum_{n=0}^{\infty} \frac{1}{n!}\int d^4x
    d^2\theta\,
W_{n}(\phi) f_{\star}^n~,
\end{equation}
here $W_{n}$ are the expansion coefficients taken at the point ${\phi}$
and the superfield $f$
is defines as
\begin{equation}\label{f} f= \Phi(y,\theta)-\phi(y)= \theta \kappa
-\theta^2 F
\end{equation}
and $f_{\star}^n=\underbrace{f\star f\star \cdots\star f}_n$. Our
first aim is calculation of this star-product.

We begin with consideration of several first orders
\begin{equation}\label{fn}
\begin{array}{rl} f_{\star}^2 =&\theta\kappa
\theta\kappa -(\kappa_\alpha +\theta_\alpha
 F)C^{\alpha\beta}(\kappa_\beta +\theta_\beta F) +\lambda F^2
=-2\theta^2 \kappa^2 +\lambda F^2~,\\ f_{\star}^3 =&\lambda F^2 \cdot
f(\theta) +2\lambda F \kappa^2~,\\ f_{\star}^4 = &-4\theta^2 \kappa^2
\lambda F^2 +\lambda^2 F^4~,\\ f_{\star}^5= &\lambda^2F^4 \cdot f
+4\lambda^2 F^3 \kappa^2~,\\ f_{\star}^6 =&-6 \theta^2 \kappa^2
\lambda^2 F^4 +\lambda^3 F^6~,\\ f_{\star}^7=&\lambda^3 F^6 \cdot f
+6\lambda^3 F^5 \kappa^2~,\\ \ldots & \ldots~,
\end{array}
\end{equation}
Then using induction ones arrive at the following expression for even
orders $2m$
\begin{equation}\label{f:even} f_{\star}^{2m}= -2m\theta^2
\kappa^2(\lambda F^2)^{m-1} +(\lambda F^2)^m~,
\end{equation}
and for odd orders $2m+1$
\begin{equation}\label{f:odd}
f_{\star}^{2m+1}= (\lambda F^2)^m  f(\theta) +2m \kappa^2 (\lambda^m
F^{2m-1})~.
\end{equation}
Collecting from (\ref{f:even}, \ref{f:odd})
terms with $\theta^2$, which will survive after
integration over chiral coordinates we obtain the component form of the
chiral superpotential
\begin{equation}\label{comp:chir}
\begin{array}{rl}
\displaystyle\int d^4x d^2\theta\,
W(\Phi)_{\star}=&\displaystyle\int d^4 x
\,\sum_{n=0}^{\infty}\frac{1}{(2n)!} W_{2n}(\phi)\cdot 2n
\kappa^2(\lambda
F^2 )^{n-1} \\
+ & \displaystyle  \int d^4 x
\,\sum_{n=0}^{\infty}\frac{1}{(2n+1)!}W_{2n+1}(\phi) \cdot
\lambda^n F^{2n+1}~.
\end{array}
\end{equation}

The antichiral superpotential expansion around the scalar field
$\bar\phi$ is defined as a series
\begin{equation}\label{antichir}
    \int d^4xd^2\bar{\theta}\,
\bar{W}_{\star}(\bar{\Phi}(\bar{y}, \bar{\theta}))=
\sum_{\bar{n}=0}^{\infty} \frac{1}{\bar{n}!}\int d^4x
    d^2\bar{\theta}\,
\bar{W}_{\bar{n}}(\bar\phi) \bar{f}_{\star}^{\bar{n}}~,
\end{equation}
here $\bar{f}_{\star}^n = \underbrace{\bar{f}\star \bar{f}\star
\ldots \star \bar{f}}_{n}$ is a product which will be obtained
below and $\bar{W}_{\bar{n}}=\frac{\partial^n
\bar{W}(\bar{\Phi})}{\partial \bar{\Phi}^n}$ -- expansion
coefficients.

The antichiral superfield being written in chiral coordinates looks as follows
\begin{equation}
    \bar{\Phi}(\bar{y},\bar{\theta})=\bar{\Phi}(y,\bar{\theta})
-i\theta^\alpha (\partial_{\alpha\dot\alpha}
\bar{\Phi}(y,\bar{\theta}))\bar{\theta}^{\dot\alpha}+\theta^2\bar{\theta}^2
\Box \bar{\Phi}(y,\bar{\theta})~,
\end{equation}
where $\Box=\frac{1}{2}\partial^{\alpha\dot\alpha}\partial_{\alpha\dot\alpha}$.
That means
\begin{equation}\label{barf}
\begin{array}{rl}
    \bar{f}=&\bar{\Phi}(\bar{y},\bar{\theta})-\bar{\phi}(y)\\
    =&\bar{\theta}^{\dot\alpha}\bar{\kappa}_{\dot\alpha}(y)
-\bar{\theta}^2 \bar{F}(y)
-i\theta^{\alpha}(\partial_{\alpha\dot\alpha}\bar{\phi}(y))\bar{\theta}^{\dot\alpha}
+i\theta^\alpha
\partial_{\alpha\dot\alpha}\bar{\kappa}^{\dot\alpha}(y)
\bar{\theta}^2 +\theta^2\bar{\theta}^2 \Box \bar{\phi}(y)~.
\end{array}
\end{equation}
Taking into account further integration over chiral coordinates $d^2\bar{\theta}$ we will consider
only components proportional to $\bar{\theta}^2$
\begin{equation}
\begin{array}{l}
\bar{f}|_{\bar{\theta}^2} = -\bar{F} +i\theta^\alpha
\partial_{\alpha\dot\alpha} \bar{\kappa}^{\dot\alpha} +\theta^2 \Box \bar{\phi}~,\\
\bar{f}_{\star}^2|_{\bar{\theta}^2} =  -2\bar{\kappa}^2 +2i
\theta^\alpha (\partial_{\alpha\dot\alpha}
\bar{\phi})\bar{\kappa}^{\dot\alpha} + \theta^2
\partial^{\alpha\dot\alpha}\bar{\phi} \partial_{\alpha\dot\alpha}\bar{\phi} +
 C^{\alpha\beta} \partial^{\dot\alpha}_\alpha \bar{\phi}
\partial_{\beta\dot\alpha} \bar{\phi}~.
    \end{array}
\end{equation}
The last term in the second line is equal to zero
due to a property
$\partial^{\dot\alpha}_{(\alpha} \bar{\phi}
\partial_{\beta)\dot\alpha} \bar{\phi}=-\partial_{\dot\alpha(\alpha} \bar{\phi}
\partial_{\beta)}^{\dot\alpha} \bar{\phi}\equiv 0$ and can be dropped. Therefore deformation
doesn't affect on antichiral sector $\bar{f} \star \bar{f}=\bar{f}
\cdot \bar{f}$ in accordance with Ref. \cite{seib}. All other
orders is equal to zero, because each $\bar{f}$ contains
$\bar{\theta}$ and $\bar{f}^2\sim \bar{\theta}^2$, i.e.
$\bar{f}_{\star}^{n}=0, \; n>2$.  Rewriting Eq.(\ref{antichir}) as
an integral over the whole superspace
\begin{equation}\label{comp:antichir0} \int d^4 xd^2\bar{\theta}\,
    \bar{W}_{\star}=-\int d^4 x d^2\bar{\theta}d^2\theta \theta^2
\bar{W}_{\star}(\bar{\phi}+ \bar{f})~, \end{equation} where
$\bar{\phi}= \bar{\phi}(y)$ and using the property $-\int
d^2\bar{\theta}d^2\theta \theta^2\bar{W}(\bar{\phi}(x))=0$ one can
finally write a component form for the antichiral superpotential
\begin{equation}\label{comp:antichir}
\begin{array}{rl}
     \displaystyle\int d^4 xd^2\bar{\theta}\, \bar{W}_{\star}= &\displaystyle
     -\int d^2\bar{\theta}d^2\theta \theta^2 \bar{\theta}^2
\left(-\bar{W}_{\bar{1}}(\bar{\phi}) \bar{F}
+\frac{1}{2}\bar{W}_{\bar{2}}(\bar{\phi})(-2\bar{\kappa}^2)\right)\\
&\displaystyle= \int d^4x \,\left(\bar{W}_{\bar{1}}(\bar{\phi})
\bar{F} + \bar{W}_{\bar{2}}(\bar{\phi})\bar{\kappa}^2\right)~,
\end{array}
\end{equation}
where the expansion coefficients $\bar{W}_{\bar{1}},
\bar{W}_{\bar{2}}$ were defined above. As one can see the great
difference between forms of chiral and antichiral superpotentials
appears. Obviously the action doesn't have Hermiticity properties.

The most nontrivial calculation is related to the K\"ahler
potential decomposition. We will suppose that its expansion is
fully symmetrical in powers of $\bar{f}$ and $f$, i.e.
\begin{equation}
    K(\Phi, \bar{\Phi})_{\star} = \sum_{m=0}^{\infty}{1\over m!}
    \left(f
\frac{\partial}{\partial \Phi}+ \bar{f} \frac{\partial}{\partial
\bar{\Phi}}\right)^m_{\star} K(\Phi, \bar{\Phi})|_{\Phi= \phi, \;
\bar{\Phi}= \bar{\phi}}~,
\end{equation}
Such kind of ordering leads to the following expansion
\begin{equation}\label{k:exp}
\begin{array}{rl}
   \displaystyle K(\Phi, \bar{\Phi})_{\star}=&K(\phi, \bar{\phi}) +K_{1} f
+K_{\bar{1}} \bar{f}\\
+ &\frac{1}{2}K_{2}f\star f +\frac{1}{2}K_{\bar{2}}\bar{f}\star
\bar{f} + \frac{1}{2} K_{1\bar{1}}(f \star \bar{f}+\bar{f} \star f
)
\\
+ &\frac{1}{3!}K_{3}f \star f \star f
 \\
+ &\frac{1}{3!}K_{2\bar{1}}(f\star f\star\bar{f} + f\star
\bar{f}\star f + \bar{f}\star f\star f) \\
+ &\frac{1}{3!}K_{1\bar{2}}(f\star \bar{f}\star\bar{f} +
\bar{f}\star \bar{f}\star f + \bar{f}\star
f\star \bar{f})+\ldots  \\
= &\displaystyle \sum_{n} K_{n}f^{n}_{\star} + \sum_{\bar{n}}
K_{\bar{n}}\bar{f}^{\bar{n}}_{\star} + \sum_{n,\bar{n}}
K_{n\bar{n}}[f^{n}_{\star}\star\bar{f}^{\bar{n}}_{\star}]~,
\end{array}
\end{equation}
where $[\bar{f}^{\bar{n}}\star f^n]$ is a fully symmetrized
star-product including all possible permutations. For instance,
$[f\star f\star\bar{f}]= f\star f\star\bar{f} + f\star
\bar{f}\star f + \bar{f}\star f\star f$. From (\ref{fn}) follows
that unmixed products like $f_{\star}^{n}$ for any $n$ will not
give contribution to the K\"ahler potential because they do not
contain factor $\bar{\theta}^2$ we need for further
integration over $\int d^2\bar{\theta}$. Unmixed star products
$\bar{f}_{\star}^{\bar{n}}$ for $n=3$ and higher will vanish and
hence, do not contribute to the action. Thus, we should study the
star product $[\bar{f}^{\bar{n}}\star f^m]$ of arbitrary integer
$m$ with $\bar{n}=1,2$. Indeed, let us consider the possible mixed
star-products in the expansion (\ref{k:exp}). Using the
superfields $f$ and $\bar{f}$ (\ref{f}, \ref{barf}), we obtain for the
first order mixed product the following expression
\begin{equation}
\begin{array}{l}
    f \star \bar{f}=\theta \kappa(\bar{\theta}\bar{\kappa}
-\bar{\theta}^2 \bar{F}-i\theta^\alpha
\bar{\theta}^{\dot\alpha}\partial_{\alpha\dot\alpha}\bar{\phi}
+i\theta^\alpha
\partial_{\alpha\dot\alpha}\bar{\kappa}^{\dot\alpha}\bar{\theta}^2)
-\theta^2 F(\bar{\theta}\bar{\kappa} -\bar{\theta}^2 \bar{F})
-\lambda F\bar{\theta}^2\Box \bar{\phi} \\
-C^{\alpha\beta}\kappa_\alpha(-i\partial_{\beta\dot\alpha}
\bar{\theta}^{\dot\alpha}\bar{\phi}+i\partial_{\beta\dot\alpha}\bar{\kappa}^{\dot\alpha}
\bar{\theta}^2 +\theta_\beta \bar{\theta}^2 \Box
\bar{\phi})-C^{\alpha\beta} F
\theta_\alpha(i\partial_{\beta\dot\alpha}
\bar{\theta}^{\dot\alpha}\bar{\phi}-i\partial_{\beta\dot\alpha}\bar{\kappa}^{\dot\alpha}
\bar{\theta}^2)~.
\end{array}
\end{equation}
After symmetrization most of the terms here will disappear and ones get
\begin{equation}
    \frac{1}{2}[f \star \bar{f} +\bar{f} \star f]=\theta
\kappa(\bar{\theta}\bar{\kappa} -\bar{\theta}^2
\bar{F}-i\theta^\alpha
\bar{\theta}^{\dot\alpha}\partial_{\alpha\dot\alpha}\bar{\phi}
+i\theta^\alpha
\partial_{\alpha\dot\alpha}\bar{\kappa}^{\dot\alpha}\bar{\theta}^2)
-\theta^2 F(\bar{\theta}\bar{\kappa} -\bar{\theta}^2 \bar{F})
-\lambda F\bar{\theta}^2\Box \bar{\phi}~.
\end{equation}
Further, because we are interested in terms with coefficient $\bar{\theta}^2$,
we concentrate only on these terms.
Third order mixed star-product looks like
\begin{equation}\label{k21}
    \frac{1}{3!}[f\star f\star\bar{f} +
f\star \bar{f}\star f + \bar{f}\star f\star f]|_{\theta^2
\bar{\theta}^2}=\kappa^2 \bar{F} +\frac{1}{2}\lambda F^2
\Box \bar{\phi}~.
\end{equation}
For obtaining the other products ones use the property \cite{D2}
\begin{equation}
    \frac{1}{(n+1)!}[\bar{f}\star
f^n]\,|_{\theta^2\bar{\theta}^2}=\frac{1}{(n)!}\bar{f}\star f^n
|_{\theta^2\bar{\theta}^2}~,
\end{equation}
which leads to the  considerable simplifications. Direct
calculation gives us factors at the coefficients $K_{\bar{1}n}$:
\begin{equation}\label{kn1}
\begin{array}{l}
    \bar{f} \star f_{\star}^{2n}|_{\theta^2\bar{\theta}^2}= 2n\kappa^2
(\lambda F^2)^{n-1}\bar{F} + (\lambda F^2)^n \Box \bar{\phi}~,\\
\bar{f} \star
f^{(2n+1)}_{\star}|_{\theta^2\bar{\theta}^2}=\lambda^n
F^{2n+1}\bar{F} -i\kappa^\alpha
\partial_{\alpha\dot\alpha}\bar{\kappa}^{\dot\alpha} \lambda^n
F^{2n} +2n\kappa^2\lambda^n F^{2n-1} \Box \bar{\phi}~.
\end{array}
\end{equation}
Next, we compute factors at the coefficients $K_{\bar{2}n}$ by the
same way
\begin{equation}\label{kn2}
    \begin{array}{l}
f_{\star}^{2n}\star \bar{f}_{\star}^2 |_{\theta^2\bar{\theta}^2}=
2\kappa^2 \bar{\kappa}^2 2n(\lambda F^2)^{n-1} +\lambda^n
F^{2n}\partial^{\alpha\dot\alpha}\bar{\phi}\partial_{\alpha\dot\alpha}\bar{\phi}~,\\
f_{\star}^{2n+1}\star \bar{f}_{\star}^2
|_{\theta^2\bar{\theta}^2}=-(\lambda F^2)^n 2i \kappa^\alpha
\bar{\kappa}^{\dot\alpha} (\partial_{\alpha\dot\alpha} \bar{\phi})
+2\bar{\kappa}^2 \lambda^n F^{2n+1}+2n \kappa^2 \lambda^n F^{2n-1}
\partial^{\alpha\dot\alpha} \bar{\phi}\partial_{\alpha\dot\alpha}
\bar{\phi}~.
    \end{array}
\end{equation}

Using (\ref{comp:chir}, \ref{comp:antichir}, \ref{kn1}, \ref{kn2})
we write the full Lagrangian in component form for the ${\cal
N}={1\over 2}$ nonanticommutative generic chiral superfield model
(\ref{genact}) as a infinite series expansion in the parameter
deformation
\begin{equation}\label{comp:genact}
\begin{array}{rl}
{\cal L}_{\star} = &K(\Phi, \bar{\Phi})_{\star}|_{\theta^2\bar{\theta}^2}+
W(\Phi)_{\star}|_{\theta^2}+\bar{W}(\bar{\Phi})_{\star}|_{\bar{\theta}^2}\\
=&\displaystyle\bar{W}_{\bar{1}} \bar{F} + \bar{W}_{\bar{2}}
\bar{\kappa}^2+ \sum_{n=0}^{\infty} \frac{\lambda^n
F^{2n}}{(2n+1)!}\left(W_{2n+2}\kappa^2+W_{2n+1} F\right)\\
+&\displaystyle\bar{F}\sum_{n=0}^{\infty}\frac{\lambda^n
F^{2n}}{(2n+1)!}\left(K_{\bar{1}(2n+2)} \kappa^2
+K_{\bar{1}(2n+1)}
F\right)\\
+&\displaystyle\Box \bar{\phi}\sum_{n=0}^{\infty}\frac{\lambda^n
F^{2n-1}}{(2n)!}\left(\frac{2n}{2n+1} K_{\bar{1}(2n+1)}\kappa^2
+K_{\bar{1}(2n)} F\right)\\
+&\displaystyle\sum_{n=0}^{\infty}\frac{\lambda^n
F^{2n}}{(2n+1)!}K_{\bar{1}(2n+1)}(i\kappa^\alpha
\partial_{\alpha}^{\dot\alpha}\bar{\kappa}_{\dot\alpha})\\
+&\displaystyle\frac{1}{2}\partial^{\alpha\dot\alpha}\bar{\phi}
\partial_{\alpha\dot\alpha}\bar{\phi}\sum_{n=0}^{\infty}\frac{\lambda^n F^{2n-1}}{(2n)!}
\left(K_{\bar{2}(2n+1)}\frac{2n}{2n+1}\kappa^2 +K_{\bar{2}(2n)} F\right)\\
+&\displaystyle\sum_{n=0}^{\infty} \frac{\lambda^n
F^{2n+1}}{(2n+1)!}K_{\bar{2}(2n+1)}\bar{\kappa}^2 \\
+&\displaystyle\sum_{n=0}^{\infty}\left[
\frac{1}{(2n)!}K_{\bar{2}(2n)}\left(2n\kappa^2 \bar{\kappa}^2
(\lambda F^2)^{n-1} \right)\right.\\
+&\displaystyle\left.\frac{1}{(2n+1)!}K_{\bar{2}(2n+1)}\left((\lambda
F^2)^n i \kappa^\alpha (\partial_{\alpha}^{\dot\alpha}
\bar{\phi})\bar{\kappa}_{\dot\alpha} \right)\right]~,
\end{array}
\end{equation}
where all coefficients are calculated at the point $\phi$, i.e.
$W_{n}= W_{n}(\phi)$, $\bar{W}_{\bar{n}}=
\bar{W}_{\bar{n}}(\bar\phi)$, $K_{n\bar{n}}=K_{n\bar{n}}(\phi,
\bar{\phi})$. The Lagrangian (\ref{comp:genact}) can be written as
a sum
\begin{equation}\label{x}
{\cal L}_{\star}={\cal L}+\Delta {\cal L}(\lambda)~,
\end{equation}
here ${\cal L}$ is the component Lagrangian for the generic chiral
superfield model in ${\cal N}=1$ superspace with the action (see
e.g. Ref. \cite{idea})
\begin{equation}\label{spoint} S[\bar{\Phi}, \Phi] =
    \int d^4x d^4\theta\, K(\bar{\Phi}, \Phi) + \int d^4x
    d^2\theta\,W(\Phi) + \int d^4x d^2\bar\theta\,\bar{W}(\bar\Phi)~.
\end{equation}
 Further we will explore some properties of the model
(\ref{spoint}). In particular, being expanded around the bosonic
fields $\phi, \bar\phi$, the component form for the Lagrangian
(\ref{spoint}) is written as
\begin{equation}\label{comp:smod}
\begin{array}{l}
\displaystyle {\cal L} = \left(-g\frac{1}{2}
\partial^{\alpha\dot\alpha} \phi
\partial_{\alpha\dot\alpha} \bar{\phi}
 +i g\kappa^\alpha
\partial_{\alpha}^{\dot\alpha} \bar{\kappa}_{\dot\alpha}- K_{1\bar{2}}i\kappa^\alpha
\bar{\kappa}^{\dot\alpha}\partial_{\alpha\dot\alpha}\bar{\phi}\right.\\
\displaystyle \phantom{\int}+ gF \bar{F}+K_{2 \bar{1}}\kappa^2 \bar{F}
+K_{1\bar{2}}\bar{\kappa}^2 F+W_{1}F + \bar{W}_{\bar{1}}\bar{F} \\
\displaystyle\left.\phantom{1\over 2} + W_{2}\kappa^2 +
\bar{W}_{\bar{2}}\bar{\kappa}^2+K_{2\bar{2}}\kappa^2
\bar{\kappa}^2 \right)~,
\end{array}
\end{equation}
where we introduced the K\"ahlerian metrics $g=
K_{1\bar{1}}(\bar{\phi}, \phi)=\partial^2 K (\bar{\phi},
\phi)/\partial\phi \partial\bar{\phi}$. Such a form can be
directly obtained from (\ref{comp:genact}) as a coefficient at
$n=0$.

Equations of motion on auxiliary fields $F, \bar{F}$ in
(\ref{comp:smod}) have the solutions
\begin{equation}\label{fpert0}
    F=-g^{-1}(\bar{W}_{\bar{1}}+K_{2\bar{1}}\kappa^2)~, \quad
\bar{F}=-g^{-1}(W_{1}+K_{1\bar{2}}\bar{\kappa}^2)~.
\end{equation}
These solutions can be used to eliminate the auxiliary fields from
the Lagrangian (\ref{comp:smod}). It gives to
\begin{equation}\label{comp:smod:ff}
\begin{array}{l}
\displaystyle {\cal L} = -g\frac{1}{2}
\partial^{\alpha\dot\alpha} \phi
\partial_{\alpha\dot\alpha} \bar{\phi}
 +i g\kappa^\alpha\partial_{\alpha}^{\dot\alpha} \bar{\kappa}_{\dot\alpha}- K_{1\bar{2}}i\kappa^\alpha
\bar{\kappa}^{\dot\alpha}\partial_{\alpha\dot\alpha}\bar{\phi}\\
\displaystyle \phantom{\int}
- g^{-1}K_{2\bar{1}}W_{1}\kappa^2 -
g^{-1}K_{2\bar{1}}K_{1\bar{2}}\bar{\kappa}^2\kappa^2-
g^{-1}K_{1\bar{2}}\bar{W}_{\bar{1}}\bar{\kappa}^2-g^{-1}W_{1}\bar{W}_{\bar{1}} \\
\displaystyle\phantom{1\over 2} + W_{2}\kappa^2 +
\bar{W}_{\bar{2}}\bar{\kappa}^2+K_{2\bar{2}}\kappa^2
\bar{\kappa}^2 ~,
\end{array}
\end{equation}
where the first line is kinetic terms while the second and third
line present the potential for the model (\ref{spoint})
\begin{equation}\label{u}
\begin{array}{c}
U =  g^{-1}K_{2\bar{1}}W_{1}\kappa^2
+g^{-1}K_{2\bar{1}}K_{1\bar{2}}\bar{\kappa}^2\kappa^2
+ g^{-1}K_{1\bar{2}}\bar{W}_{\bar{1}}\bar{\kappa}^2 \\
-g^{-1}W_{1}\bar{W}_{\bar{1}}
- W_{2}\kappa^2
- \bar{W}_{\bar{2}}\bar{\kappa}^2
-K_{2\bar{2}}\kappa^2 \bar{\kappa}^2~.
\end{array}
\end{equation}
Since the general chiral superfield model on ${\cal N}={1\over 2}$
superspace includes the action (\ref{spoint}) we expect that the
potential (\ref{u}) for the model under consideration should be
modified by  dependent on deformation parameter $\lambda$ terms.

The Lagrangian (\ref{comp:smod}) presents all terms independent on
$\lambda$ (which survive at $\lambda =0$ in (\ref{comp:genact})),
while the new term $\Delta{\cal L}$ is conditioned by the
superspace deformation and depends on the parameter $\lambda$
\begin{equation}\label{comp:genact:d}
\begin{array}{rl}
\Delta{\cal L}(\lambda)=&\displaystyle\sum_{n=1}^{\infty}
\frac{\lambda^n
F^{2n}}{(2n+1)!}\left(W_{2n+2}\kappa^2+W_{2n+1} F\right)\\
+&\displaystyle\bar{F}\sum_{n=1}^{\infty}\frac{\lambda^n
F^{2n}}{(2n+1)!}\left(K_{\bar{1}(2n+2)} \kappa^2
+K_{\bar{1}(2n+1)}
F\right)\\
+&\displaystyle\Box \bar{\phi}\sum_{n=1}^{\infty}\frac{\lambda^n
F^{2n-1}}{(2n)!}\left(\frac{2n}{2n+1} K_{\bar{1}(2n+1)}\kappa^2
+K_{\bar{1}(2n)} F\right) \\
+&\displaystyle\sum_{n=1}^{\infty}\frac{\lambda^n
F^{2n}}{(2n+1)!}K_{\bar{1}(2n+1)}(i\kappa^\alpha
\partial_{\alpha}^{\dot\alpha}\bar{\kappa}_{\dot\alpha})\\
+&\displaystyle\frac{1}{2}\partial^{\alpha\dot\alpha}\bar{\phi}
\partial_{\alpha\dot\alpha}\bar{\phi}\sum_{n=1}^{\infty}\frac{\lambda^n F^{2n-1}}{(2n)!}
\left(K_{\bar{2}(2n+1)}\frac{2n}{2n+1}\kappa^2 +K_{\bar{2}(2n)} F\right)\\
+&\displaystyle\sum_{n=1}^{\infty} \frac{\lambda^n
F^{2n+1}}{(2n+1)!}K_{\bar{2}(2n+1)}\bar{\kappa}^2 \\
+&\displaystyle\sum_{n=1}^{\infty}\left[
\frac{1}{(2n)!}K_{\bar{2}(2n)}\left(2n\kappa^2 \bar{\kappa}^2
(\lambda F^2)^{n-1} \right)\right.\\
+&\displaystyle\left.\frac{1}{(2n+1)!}K_{\bar{2}(2n+1)}\left((\lambda
F^2)^n i \kappa^\alpha (\partial_{\alpha}^{\dot\alpha}
\bar{\phi})\bar{\kappa}_{\dot\alpha} \right)\right]~,
\end{array}
\end{equation}
The relations (\ref{x}, \ref{comp:smod}, \ref{comp:genact:d}) give
the component structure for the considered deformed generic chiral
superfield model.

The obtained representation for the action (\ref{comp:genact}) is
complicated and inaccessible even in the classical domain. Now we
show that the infinite series (\ref{comp:genact}) can be resummed
in a compact expression similar to the standard Zumino's
Lagrangian \cite{Zum} with the deformed K\"ahler potential and the
chiral superpotential plus a finite number of higher dimensional
terms with field-dependent couplings. In the analogy with the
trick used in the papers \cite{ketov, alva} we introduce "fuzzy
fields" controlled by the auxiliary fields $\phi +\tau
\sqrt{\lambda} F$ on interval $-1 \leq \tau \leq 1$:
\begin{eqnarray}
&\displaystyle{\cal W}^{(0)}(\phi, F) =\frac{1}{2}\int^1_{-1} d\tau W(\phi +\tau
\xi)~, \quad \xi= \sqrt{\lambda} F~, \nonumber\\
& \displaystyle{\cal K}^{(0)}(\phi,
F,\bar{\phi})=\frac{1}{2}\int^1_{-1} d\tau K(\phi +\tau \xi,
\bar{\phi})~, \nonumber\\
& \displaystyle{\cal K}^{(1)}(\phi, F,\bar{\phi})=\frac{1}{2}\int^1_{-1}
d\tau \tau K(\phi +\tau \xi,\bar{\phi})~,\nonumber\\
&\displaystyle{\cal K}^{(-1)}(\phi,
F,\bar{\phi})=\frac{1}{2}\int^1_{-1} d\tau
\frac{\partial}{\partial \tau}(\tau \cdot K(\phi +\tau
\xi,\bar{\phi}))~.
\end{eqnarray}
Then (\ref{comp:genact}) can be rewritten in a compact form:
\begin{eqnarray}
{\cal L}_\star =\bar{W}_{\bar{1}} \bar{F} + \bar{W}_{\bar{2}}
\bar{\kappa}^2 +F {\cal W}^{(0)}_{1} +\kappa^2 {\cal W}^{(0)}_{2}
+(\bar{F}F +i\kappa^\alpha
\partial^{\dot\alpha}_\alpha \bar{\kappa}_{\dot\alpha}){\cal K}^{(0)}_{1\bar{1}} +\kappa^2
\bar{F}{\cal K}^{(0)}_{2\bar{1}}\nonumber\\
+\Box \bar{\phi}{\cal K}^{(-1)}_{\bar{1}} +\sqrt{\lambda}\kappa^2
\Box \bar{\phi}{\cal K}^{(1)}_{2\bar{1}}
+\frac{1}{2}\partial^{\alpha\dot\alpha}\bar{\phi}
\partial_{\alpha\dot\alpha}\bar{\phi} {\cal K}^{(-1)}_{\bar{2}} +
\bar{\kappa}^2 F{\cal K}^{(0)}_{1\bar{2}}\label{k01}\\
+i\kappa^\alpha(\partial^{\dot\alpha}_\alpha
\bar{\phi})\bar{\kappa}_{\dot\alpha}{\cal K}^{(0)}_{1\bar{2}}
+\sqrt{\lambda}\kappa^2\frac{1}{2}\partial^{\alpha\dot\alpha}\bar{\phi}
\partial_{\alpha\dot\alpha}\bar{\phi} {\cal K}^{(1)}_{2\bar{2}}
+\kappa^2\bar{\kappa}^2 {\cal K}^{(0)}_{2\bar{2}}~.\nonumber
\end{eqnarray}
It is quite remarkable that the deformation encoded by new
geometric quantities which look like the "metric" ${\cal
K}^{(0)}_{1\bar{1}}$, "connection" ${\cal K}^{(0)}_{2\bar{1}}$ and
the "curvature" ${\cal K}^{(0)}_{2\bar{2}}$ in the smearing target
space. But there is no any certainty that this quantities are
really consistent among themselves and correspond to some
geometrical structure of the target space manifold.

It is easy to see that the first and the third terms in the second
line of (\ref{k01}) up to space-time derivatives of the auxiliary
field $F$ can be written as
 $-\frac{1}{2}\partial^{\alpha\dot\alpha}\phi
\partial_{\alpha\dot\alpha}\bar{\phi}{\cal K}^{(-1)}_{1\bar{1}}$.
Second term in the second line plus second term in the third line
is
$-\sqrt{\lambda}\kappa^2\frac{1}{2}\partial^{\alpha\dot\alpha}\phi
\partial_{\alpha\dot\alpha}\bar{\phi}{\cal K}^{(1)}_{3\bar{1}}
-\sqrt{\lambda}(\partial^{\alpha\dot\alpha}\kappa^2)\frac{1}{2}
\partial_{\alpha\dot\alpha}\bar{\phi}{\cal K}^{(1)}_{2\bar{1}}$.
We can rewrite (\ref{k01}) in the canonical form with a proper
kinetic term for the scalars $\partial^{\alpha\dot\alpha}\phi
\partial_{\alpha\dot\alpha}\bar{\phi}{\cal K}^{(0)}_{1\bar{1}}$
but, due to the extra dependence of ${\cal K}^{(0)}(\phi, F,
\bar{\phi})$ of the auxiliary field $F$, there will be new terms
containing one derivative of the auxiliary field
$\partial^{\alpha\dot\alpha} F$. At the limit $\lambda\rightarrow
0$ this terms will vanish. This is the great difference between
(\ref{comp:smod}) and (\ref{k01}).

Taking into account properties $\frac{\partial}{\partial F}{\cal
K}^{(0)} =\sqrt{\lambda}\frac{\partial}{\partial \phi}{\cal
K}^{(1)}$ and ${\cal K}^{(-1)} = {\cal K}^{(0)} +
F\frac{\partial}{\partial F} {\cal K}^{(0)}$ one can note that Eq.
(\ref{k01}) has the structure similar to Eq.(\ref{comp:smod}),
where the quantity ${\cal K}^{(0)}_{1\bar{1}}$ may be considered
as a deformed metric dependent on the auxiliary field $F$.
Combining terms one can rewrite the expression (\ref{k01}) via
single function ${\cal K}^{(0)}$ as follows:
\begin{eqnarray}
{\cal L}_\star =\bar{W}_{\bar{1}} \bar{F} + \bar{W}_{\bar{2}}
\bar{\kappa}^2 +F {\cal W}^{(0)}_{1} +\kappa^2 {\cal W}^{(0)}_{2}
+(\bar{F}F +i\kappa^\alpha
\partial^{\dot\alpha}_\alpha \bar{\kappa}_{\dot\alpha})
{\cal K}^{(0)}_{1\bar{1}} +\kappa^2
\bar{F}{\cal K}^{(0)}_{2\bar{1}}\nonumber\\
+\Box \bar{\phi}{\cal K}^{(0)}_{\bar{1}} +F\Box
\bar{\phi}\partial_F {\cal K}^{(0)}_{\bar{1}} +\kappa^2 \Box
\bar{\phi}\partial_F {\cal K}^{(0)}_{1\bar{1}}
+\frac{1}{2}\partial^{\alpha\dot\alpha}\bar{\phi}
\partial_{\alpha\dot\alpha}\bar{\phi} {\cal
K}^{(0)}_{\bar{2}}+\frac{1}{2}F\partial^{\alpha\dot\alpha}\bar{\phi}
\partial_{\alpha\dot\alpha}\bar{\phi} \partial_F{\cal
K}^{(0)}_{\bar{2}}\\
+\bar{\kappa}^2 F{\cal K}^{(0)}_{1\bar{2}}
+i\kappa^\alpha(\partial^{\dot\alpha}_\alpha
\bar{\phi})\bar{\kappa}_{\dot\alpha}{\cal K}^{(0)}_{1\bar{2}}
+\kappa^2\frac{1}{2}\partial^{\alpha\dot\alpha}\bar{\phi}
\partial_{\alpha\dot\alpha}\bar{\phi} \partial_F{\cal K}^{(0)}_{1\bar{2}}
+\kappa^2\bar{\kappa}^2 {\cal K}^{(0)}_{2\bar{2}}~,\nonumber
\end{eqnarray}
where $\partial_{F}= \partial/\partial F$. Comparison with Eq.
(\ref{comp:smod}) shows the terms which spoil the K\"ahler
structure of the Lagrangian
\begin{equation}
{\cal L}_\star = {\cal L}(W\rightarrow {\cal W}^{(0)},\; K\rightarrow{\cal
K}^{(0)})\end{equation}
$$+\Box \bar{\phi} F\partial_{F}{\cal
K}^{(0)}_{\bar{1}} +\frac{1}{2}F
\partial^{\alpha\dot\alpha}\bar{\phi}
\partial_{\alpha\dot\alpha}\bar{\phi}\partial_{F}{\cal
K}^{(0)}_{\bar{2}} +\kappa^2 \Box\bar{\phi}\partial_{F}{\cal
K}^{(0)}_{1\bar{1}}
+\kappa^2\frac{1}{2}\partial^{\alpha\dot\alpha}\bar{\phi}
\partial_{\alpha\dot\alpha}\bar{\phi}\partial_{F}{\cal
K}^{(0)}_{1\bar{2}}~.
$$
This is  quite natural because though the initial Lagrangian
(\ref{genact}) has the K\"ahlerian form the ${\cal N}= {1\over 2}$
superspace is not a K\"ahlerian manifold ($\bar{\theta}$ is not
the complex conjugate of $\theta$). This property was firstly
noted in the recent work \cite{alva} for ${\cal N}=2$, $D=2$
nonanticommutative sigma model.

Now consider generic nonanticommuting supersymmetric sigma-model
(i.e. the model without superpotential $W$ but with arbitrary
Kahlerian potential $K$). It was shown in Ref. \cite{chandr} that
for $D=2$, ${\cal N}=2$ nonanticommuting sigma-model the component
action infinite series can be resummed to a very simple and clear
form. Let's consider such possibility for $D=4$, ${\cal
N}=\frac{1}{2}$ nonanticommuting sigma-model. In the linear
approximation on $\lambda$ the Lagrangian (\ref{comp:genact})
after introducing a new metric $\tilde{g}=g +\frac{\lambda}{6}F^2
K_{3\bar{1}}$ can be rewritten as follows
\begin{eqnarray}
{\cal L}_{\star}= -\frac{1}{2}\partial^{\alpha\dot\alpha}\phi
\partial_{\alpha\dot\alpha}\bar{\phi}(g+ \frac{\lambda}{2}F^2 K_{3\bar{1}} +
\frac{\lambda}{3}F\kappa^2 K_{4\bar{1}})+(F\bar{F} +i\kappa^\alpha
\partial^{\dot\alpha}_\alpha \bar{\kappa}_{\dot\alpha} )\tilde{g}\nonumber\\
+ i\kappa^\alpha (\partial^{\dot\alpha}_\alpha
\bar{\phi})\bar{\kappa}_{\dot\alpha} \tilde{g}_{\bar{1}}
+\bar{F}\kappa^2 \tilde{g}_1
+{F}\bar{\kappa}^2
\tilde{g}_{\bar{1}} +\kappa^2\bar{\kappa}^2\tilde{g}_{1\bar{1}}\label{resum}\\
+ \frac{\lambda}{6}F\kappa^2 \cdot
\partial^{\alpha\dot\alpha}(\partial_{\alpha\dot\alpha} \bar{\phi}
K_{3\bar{1}} ) +\frac{1}{4}F^2 \cdot
\partial^{\alpha\dot\alpha}(\partial_{\alpha\dot\alpha} \bar{\phi}
K_{2\bar{1}}) \nonumber
\end{eqnarray}
The equation of motion for field $F$ following from this
Lagrangian is
\begin{equation}\label{aux}
F\tilde{g} +\kappa^2\tilde{g}_1 =0
\end{equation}
and at that time two last terms $ \sim \kappa^4$ that vanish. This
allows to note that the expression $(g+ \frac{\lambda}{2}F^2
K_{3\bar{1}} + \frac{\lambda}{3}F\kappa^2 K_{4\bar{1}})$ become
equal $\tilde{g}$. Thus we see that Lagrangian (\ref{resum}) in
the first order on $\lambda$ is one to one correspond to the
Zumino Lagrangian with the metric $\tilde{g}$. We point out that
such a consideration is true only $W=0$ and for a singlet
fermionic field. In accord with Ref. \cite{chandr} one can verify
that the action given by Eq. (\ref{comp:genact}) at $W=0$ and
$\bar{W}=0$ in all orders on $\lambda$ can be rewritten in the
form Eq. (\ref{resum}).

Next we discuss elimination of the auxiliary fields $F, \bar{F}$
from the component Lagrangian (\ref{comp:genact}) keep in mind the
task investigate the structure of classical vacua. The Lagrangian
(\ref{comp:genact}) is linear in $\bar{F}$ but strongly nonlinear
in $F$.  Therefore it is difficult to expect that we obtain the
exact solution on $F$ and $\bar{F}$ but we can perturbatively find
several first corrections to the scalar potential and to the
scalar - fermion interaction terms. In particular, the scalar
potential is the most important object for study the possible
vacua of the theory and a symmetry breaking mechanism studying.
Let's consider only space-time independent vacuum expectation
values for the scalar and fermionic physical fields. We suppose
that
\begin{equation}\label{fpert} F=F_0+F_1+\cdots~, \quad
\bar{F}=\bar{F}_0+\bar{F}_1+\cdots~,
\end{equation}
where $F_{0}$ and $\bar{F}_{0}$ are the solutions for auxiliary
fields equations of motion presented by the Eq. (\ref{fpert0}),
$F_{n}\sim \lambda^n$, $\bar{F}_{\bar{n}}\sim \lambda^{\bar{n}}$
are the corrections. Substituting (\ref{fpert}) into the
Lagrangian (\ref{comp:genact}) and keeping only linear in
$\lambda$ terms without derivatives we obtain first corrections to
the auxiliary fields
\begin{equation}\label{fpert1}
\begin{array}{rl} -
gF_1=&\lambda \left(\frac{1}{6}F_0^2 K_{4\bar{1}} \kappa^2
+\frac{1}{2}F_0^3 K_{3\bar{1}}
    \right)~,\\ -g\bar{F}_1=&\lambda \bar{F}_0
F_0\left(\frac{1}{3}K_{4\bar{1}}\kappa^2 +\frac{1}{2} F_0
K_{3\bar{1}} \right)+ \frac{1}{2}\lambda F_0^2
K_{3\bar{2}}\bar{\kappa}^2 \\
+&\lambda \frac{1}{3}F_0 K_{4\bar{2}}\kappa^2\bar{\kappa}^2
+\lambda F_0\left(\frac{1}{3}W_4\kappa^2 +\frac{1}{2}F_0
W_3\right)~.
\end{array}
\end{equation}

This gives us, in addition to the ordinary potential $U$ presented
by (\ref{u}), a linearly dependent on $\lambda$ correction
\begin{equation}\label{u1}
  \begin{array}{rl}
\Delta U_{1}(\lambda)=&g(F_1\bar{F}_0 +\bar{F}_1F_0)
+F_1(W_1+K_{1\bar{2}}\bar{\kappa}^2)+\bar{F}_1(\bar{W}_{\bar{1}}+K_{2\bar{1}}\kappa^2)\\
+&\frac{\lambda}{6}F_0^2 K_{4\bar{2}}\kappa^2 \bar{\kappa}^2
+\frac{\lambda}{6}F_0^2\bar{F}_0 (K_{4\bar{1}}\kappa^2 +F_0
K_{3\bar{1}})\\
+&\frac{\lambda}{6}F_0^2(W_4 \kappa^2 +F_0
W_3)+\frac{\lambda}{6}F_0^3 K_{3\bar{2}}\bar{\kappa}^2~,
\end{array}
\end{equation}
where $F_{0}, \bar{F}_{0}$ and $F_{1}, \bar{F}_{1}$ are given in
Eqs. (\ref{fpert0}, \ref{fpert1}) respectively. As a result, we
finally get that the potential $U$ given by
Eq.(\ref{comp:smod:ff}) and a series of additional terms dependent
on $\lambda$. Considering the expressions (\ref{u}, \ref{fpert1},
\ref{u1}) one can see that the full potential as a function of the
scalar fields and fermionic condensate $\langle\kappa^2\rangle$
can be as positive as negative defined depending on concrete forms
of the K\"ahlerian and chiral superpotentials. It means, in
general, that at nonvanishing $\lambda$ the potential possesses a
possibility to get a minimum, though the initial potential
(\ref{u}) has none minimum. Therefore one can expect some kind of
symmetry breaking in the model under consideration. This very
interesting aspect is out of this paper subject and deserves a
separate study.

To summarize, we have considered the supersymmetric generic chiral
superfield model on ${\cal N}={1\over 2}$ nonanticommutative
superspace. This model is given in terms of arbitrary K\"ahlerian
potential, chiral and antichiral superpotentials. We have
developed a general procedure for deriving the component structure
of the model and obtained the component action in the explicit
form as a infinite series in the nonanticommutativity parameter.
This series is summed up into compact expression using the
specific integral representations. It was shown that the
additional "deformed" part of the action allows a perturbative
translation invariant solution for the auxiliary fields equations
of motion. Leading corrections to nondeformed potential are
calculated. The results obtained can be applied to studying a wide
class of various ${\cal N}={1\over 2}$ chiral superfield models
including supersymmetric sigma-models and models with different
chiral and antichiral superpotentials.

 After we had put this work in the hep-th ArXiv, the work
\cite{ketov} appeared, where the component structure for chiral
superpotential (\ref{comp:chir}, \ref{comp:antichir}) has been
obtained. After that two papers \cite{alva} and \cite{chandr}
related to studying the component structure of $N=2$,
$D=2$ nonanticommuting sigma models appeared.

\vspace{0.2cm}

{\bf Acknowledgements} The work was supported in part by RFBR
grant, project No 03-02-16193. I.L.B is grateful to RFBR-DFG
grant, project No 04-02-04002, to DFG grant, project No 436 RUS
113/669, to LSS grant, project No 1252.2003.2 and to INTAS grant,
INTAS-03-51-6346 for partial support.


\begin{thebibliography}{000}




\bibitem{seib} N. Seiberg, {\it JHEP} {\bf 0306} (2003)
010, hep-th/0305248.



\bibitem{n11st} E. Ivanov, O. Lechtenfeld, B. Zupnik,
{\it JHEP} {\bf 0402} (2004) 012, hep-th/0308012; S. Ferrara, E.
Ivanov, O. Lechtenfeld, E. Sokatchev, B. Zupnik, {\it Nucl.Phys.}
{\bf B704} (2005) 154, hep-th/0405049; S. Ferrara and E.
Sokatchev, {\it Phys.Lett.} {\bf B579} (2004) 226, hep-th/0308021;
T. Araki, K.Ito and A. Ohtsuka, {\it JHEP} {\bf 0401} (2004) 046,
hep-th/0401012; {\it Phys.Lett.} {\bf B606} (2005) 202,
hep-th/0410203.

\bibitem{berk}N. Berkovits and N. Seiberg, {\it JHEP} {\bf 0307} (2003) 010,
 hep-th/0306226;  S. Ferrara, M. A. Lledo and O. Macia,
{\it JHEP} {\bf 0309} (2003) 068, hep-th/0307039.



\bibitem{quantn12}S. Terashima and J. T. Yee,
{\it JHEP} {\bf 0312} (2003) 053, hep-th/0306237; M. Grisaru, S.
Penati and A. Romagnoni, {\it JHEP} {\bf 0308} (2003) 003,
hep-th/0307099; A. Romagnoni, {\it JHEP} {\bf 0310} (2003) 016,
hep-th/0307209; D. Berenstein and S. J. Rey, {\it Phys.Rev.} {\bf
D68} (2003) 121701, hep-th/0308049; R. Britto, B. Feng and S. J.
Rey, {\it JHEP} {\bf 0307} (2003) 067, hep-th/0306215; R. Britto,
B. Feng, {\it Phys.Rev.Lett.} {\bf 91} (2003) 201601,
hep-th/0307165; A. Banin, I. Buchbinder, N. Pletnev, {\it JHEP}
{\bf 07} (2004) 01101, hep-th/0405063.

\bibitem{renorm}
O. Lunin, S-J Rey, {\it JHEP} {\bf 0309} (2003) 045,
hep-th/0307275; M. Alishahiha, A. Ghodsi and N. Sadooghi, {\it
Nucl.Phys.} {\bf B691} (2004) 111, hep-th/0309037; S.Penati,
A.Romagnoni, {\it JHEP} {\bf 0502} (2005) 064, hep-th/0412041.

\bibitem{D2} B. Chandrasekhar, A. Kumar,{\it JHEP} {\bf 0403} (2004) 013, hep-th/0310137.

\bibitem{D24}  B. Chandrasekhar, {\it Phys. Rev.} {\bf D70} (2004)
125003, hep-th/0408184.


\bibitem{klemm} D. Klemm, S. Penati and L. Tamassia,
{\it Class. Quant. Grav.} {\bf 20} (2003) 2905,
hep-th/0104190.

\bibitem{GSW} M.B. Green, J. Schwarz, E. Witten, Superstring Theory,
Cambridge Univ.Press, 1987, Vol 2.


\bibitem{CCE} G. Clever, M. Cveti\v{c}, J.R. Espinosa, L. Everett, P.
Langacker, {\it Nucl.Phys.} {\bf B525} (1998) 3, hep-th/9711178;
{\it Phys.Rev.} {\bf D59} (1999) 55; M. Cveti\v{c}, L. Everett, J.
Wang, {\it Nucl.Phys.} {\bf B538} (1999) 52, hep-ph/9807321; I.L.
Buchbinder, M. Cveti\v{c}, A.Yu. Petrov, {\it Nucl.Phys.} {\bf
B571} (2000) 358, hep-th/9906141; {\it Mod.Phys.Lett.}  {\bf A15}
(2000) 783, hep-th/9903243.



\bibitem{idea} I.L. Buchbinder, S.M. Kuzenko, "Ideas and Methods of
Supersymmetry and Supergravity or a Walk Through Superspace", IOP
Publishing, Bristol and Philadelphia, 1998.

\bibitem{Zum} B. Zumino, {\it Phys.Lett.} {\bf B87} (1979) 203.

\bibitem{ketov} T. Hatanaka, S.V. Ketov, Y. Kobayashi and S.
Sasaki, "Non-anticommutative Deformation of Effective Potentials
in Supersymmetric Gauge Theories", hep-th/0502026.

\bibitem{alva} Luis Alvarez-Gaume, Miguel A. Vazquez-Mozo, "On
nonanticommutative N=2 sigma-models in two dimensions",
hep-th/0503016.

\bibitem{chandr}B. Chandrasekhar, "${\cal N}=2$ $\sigma$-model Action on Non(anti)commutative
Superspace", hep-th/0503116.



\end{thebibliography}
\end{document}